\documentclass[11pt,twocolumn]{article}
\usepackage[margin=1in]{geometry}
\usepackage{amsmath,amssymb,amsfonts}
\usepackage{graphicx}
\usepackage{textcomp}
\usepackage{xcolor}
\usepackage{booktabs}
\usepackage{enumitem}
\usepackage{microtype}
\usepackage{amsthm}
\newtheorem{definition}{Definition}

\usepackage{tikz}
\usetikzlibrary{shapes,arrows,positioning}
\usepackage{hyperref}
\hypersetup{hidelinks}

\author{}
\date{}

\newcommand{\vtarget}{v_{\mathrm{target}}}

\newcommand{\taudomain}{\tau_{\mathrm{domain}}}
\newcommand{\taulow}{\tau_{\mathrm{low}}}
\newcommand{\tauhigh}{\tau_{\mathrm{high}}}

\begin{document}

\title{Topological Void Analysis: A Mathematical Framework\\
       for Systematic Technical Innovation Discovery\\
       in Knowledge Spaces}
\author{Kris Pan \\ Intel Corporation \\ \texttt{kris.pan@intel.com}}
\date{}

\twocolumn[
  \maketitle
  \begin{@twocolumnfalse}
  \begin{center}
  \begin{minipage}{0.92\textwidth}
  \textbf{Abstract.}
  Identifying where to innovate in a dense technical domain---such as
  operating systems or hardware/software co-design---is fundamentally a
  search problem in a high-dimensional knowledge space.  Existing
  approaches rely on keyword search, citation proximity, or human
  intuition, none of which formalise the notion of an \emph{unexplored
  region} that is simultaneously relevant to a target goal and absent
  from prior art.

  We present \emph{Topological Void Analysis} (TVA), a mathematical
  framework that defines \emph{topological voids} as triads $(A, B, C)$
  in a dense-sparse hybrid embedding space.  A void requires three
  conditions: (i)~both concepts $A$ and $B$ are semantically cohesive
  with domain anchor $C$; (ii)~their pairwise similarity falls within a
  calibrated marginality band---avoiding both obvious combinations and
  unrelated noise; and (iii)~they share a sparse lexical bridge while
  the geodesic midpoint on the embedding hypersphere is unoccupied.

  Applied to $\sim$140k indexed documents, TVA generates 2,128
  invention candidates across 96 targets; 90\% survive automated
  quality filtering, yielding 191~REVISE and 1~APPROVE verdict from
  four-specialist adversarial review (0.05\% end-to-end).  Two case
  studies demonstrate the framework surfaces non-obvious connective
  tissue rather than merely obvious related pairs.
  \end{minipage}
  \end{center}
  \vspace{1em}
  \end{@twocolumnfalse}
]

\section{Introduction}
\label{sec:intro}

Modern software systems, especially at the systems-software layer,
evolve through incremental invention: a developer notices a gap
between two subsystems, proposes an abstraction to bridge them, and
the community iterates toward a patch or patent.  The \emph{noticing}
step is bottlenecked by human attention and domain breadth.  A single
engineer cannot simultaneously hold in mind the locking semantics of
the Linux scheduler, the memory-ordering guarantees of eBPF JIT
output, the ELF relocation model, and the BPF verifier's type system
well enough to recognise that an IFUNC-style dispatch contract could
unify the latter three.

This paper asks: \emph{can we formalise and automate the discovery of unexplored technical gaps?}

We answer yes, and make the following contributions:
\begin{enumerate}
\item A formal definition of a \emph{topological void}
  (Section~\ref{sec:framework}): a triad $(A, B, C)$ satisfying
  domain cohesion, calibrated marginality, and sparse lexical bridge
  conditions in a hybrid dense-sparse embedding space.

\item A \emph{vacancy probing} mechanism
  (Section~\ref{sec:vacancy}) based on spherical linear interpolation
  (SLERP) that rejects pseudo-voids whose midpoint is occupied by
  existing documents.

\item An \emph{adaptive threshold calibration} procedure
  (Section~\ref{sec:calibration}) that derives domain-specific
  marginality bounds from corpus statistics, removing manual tuning.

\item A large-scale empirical evaluation
  (Section~\ref{sec:eval}) with two detailed case studies demonstrating
  that TVA surfaces non-obvious yet technically grounded innovation
  candidates.
\end{enumerate}

\section{Background and Motivation}
\label{sec:background}

\subsection{Technical Knowledge as an Embedding Space}

Pre-trained embedding models map technical documents into a
high-dimensional unit sphere~\cite{ethayarajh2019anisotropy}.  Points
close in cosine distance share semantic content; distant points are
unrelated.  A \emph{knowledge corpus} $\mathcal{K}$ is a finite set of
such points.

Innovation in this view is the act of bridging two concepts $A$,~$B$
that are not yet co-located in $\mathcal{K}$, provided their
combination is relevant to a target goal $C$.

\subsection{Geometric Convergence Across Model Scales}

A key theoretical underpinning of TVA is the
\emph{Platonic Representation Hypothesis}~\cite{huh2024platonic}:
sufficiently trained models---regardless of architecture, modality, or
scale---converge toward a common statistical geometry of their shared
training world.  Concretely, the pairwise distance structure of a
compact embedding model (BGE-M3, 1024D) and a frontier LLM are
approximately isometric: if $\mathrm{cos}(\mathrm{dense}(A),
\mathrm{dense}(B)) \approx 0$ in BGE-M3 space, the same pair tends
to be distant in the LLM's implicit representation space.

This convergence provides the theoretical basis for TVA's design.
A topological void identified in BGE-M3 space---a pair $(A,B)$ whose
geodesic midpoint is unoccupied---serves as a \emph{statistically
reliable proxy} for an underexplored region in the frontier LLM's
reasoning space.  The compact model acts as an efficient navigator;
the LLM supplies the high-resolution generative capability to fill
the identified gap.  Geometric alignment across scales, formalised via
CKA~\cite{kornblith2019similarity}, supports this proxy relationship
empirically.

\subsection{Limitations of Prior Art Search}

Conventional patent and prior-art search is keyword-driven or
citation-driven~\cite{lee2007patent, srinivasan2021patentbert}.  These
methods retrieve \emph{known} content; they do not identify
\emph{missing} content.  Knowledge-graph completion
approaches~\cite{bordes2013transe} predict missing edges but require a
pre-defined relation schema and do not model the continuous
``marginality'' notion central to patentable non-obviousness.

\subsection{The Marginality Principle}

Patent law requires that an invention be \emph{non-obvious}: too
similar to existing work fails the novelty bar; too dissimilar yields
an incoherent or non-enabling disclosure.  High-value inventions live
in a band of \emph{moderate dissimilarity}---far enough from prior art
to be novel, close enough to the domain to be useful.  This is the
informal basis for our formal marginality condition.

\section{Why Standard Approaches Fail}
\label{sec:motivation}

Three retrieval baselines were evaluated before arriving at TVA, each
failing in a characteristic way.
\textbf{Cosine top-$k$} retrieves relevant but redundant results---all
from the same well-covered subsystem, with no notion of a gap.
\textbf{MMR}~\cite{carbonell1998mmr} improves diversity but provides
no occupancy guarantee: $\sim$30\% of high-MMR pairs had a corpus
neighbour within cosine distance 0.08 of their linear midpoint,
making them false voids.
\textbf{Latent vector arithmetic}~\cite{mikolov2013linguistic}
($v_{\mathrm{bridge}} = \vtarget + (v_A - v_B)$) fails in
1024-dimensional contextualised space: anisotropy~\cite{ethayarajh2019anisotropy}
makes difference vectors nearly orthogonal to semantic axes, and the
curse of dimensionality~\cite{beyer1999nearest} strips them of
directional meaning---67\% of bridge vectors escaped the semantic
manifold entirely.
All three lack an explicit \emph{occupancy check}: they rank or
construct candidate points without verifying the region is genuinely
unoccupied.  This motivates the vacancy probe in TVA.

\section{The Topological Void Framework}
\label{sec:framework}

\subsection{Notation}

Let $\mathcal{K} = \{k_1, \ldots, k_n\}$ be a corpus of technical
documents.  Each document $k \in \mathcal{K}$ is mapped by BGE-M3
($d=1024$)~\cite{chen2024bge} to a dense unit vector
$\mathrm{dense}(k) \in S^{d-1}$ and a sparse token set
$\mathrm{sparse}(k) \subset \Sigma^*$ (top-5 lexical weights).

Let $\mathrm{cos}(u,v) = u^\top v$ for unit vectors.  Let
$\mathrm{Sparse}(k) = \{t \in \mathrm{sparse}(k) : t \notin
\mathcal{S}\}$ where $\mathcal{S}$ is a domain stop-word list
(high-frequency, low-specificity tokens such as \texttt{int},
\texttt{define}, \texttt{linux}).

\subsection{Formal Definition}

\begin{definition}[Topological Void]
\label{def:void}
Let $A, B \in \mathcal{K}$ be two corpus documents and let
$C = m(\mathrm{dense}(A),\,\mathrm{dense}(B)) \in S^{d-1}$
(Definition~\ref{def:slerp}) be their synthetic
\emph{void midpoint}---the geodesic bridge concept.
The pair $(A, B)$ forms a \emph{topological void}
with respect to a domain query vector $\vtarget \in S^{d-1}$ if
all of the following hold:

\medskip
\noindent\textbf{C1 (Domain Cohesion).}
\begin{align*}
  \mathrm{cos}(\mathrm{dense}(A), \vtarget) &> \taudomain \\
  \mathrm{cos}(\mathrm{dense}(B), \vtarget) &> \taudomain
\end{align*}

\noindent\textbf{C2 (Calibrated Marginality).}
\[
  \taulow \;\le\;
  \mathrm{cos}(\mathrm{dense}(A),\,\mathrm{dense}(B))
  \;\le\; \tauhigh
\]

\noindent\textbf{C3 (Sparse Lexical Bridge).}
\[
  \mathrm{Sparse}(A) \cap \mathrm{Sparse}(B) \;\ne\; \emptyset
\]
where $\mathrm{Sparse}(k)$ is the top-5 BGE-M3 sparse-weight set
of document $k$ after stop-word removal (as defined in
Section~\ref{sec:framework}).

\noindent\textbf{C4 (Vacancy).}
\[
  \max_{k \in \mathcal{K} \setminus \{A,B\}}
    \mathrm{cos}(\mathrm{dense}(k),\; C)
  \;<\; \theta_v
\]
where $C = m(\mathrm{dense}(A), \mathrm{dense}(B))$ is the synthetic
void midpoint (Definition~\ref{def:slerp}) and $\theta_v$ is a
vacancy threshold.
\end{definition}

Intuitively: $A$ and $B$ both point toward the target domain~(C1);
are neither trivially similar nor unrelated~(C2); share at least one
meaningful technical token as a conceptual bridge~(C3); and the
synthetic void midpoint $C$ is unoccupied by any existing
document~(C4).  The idea to be invented lives in the neighbourhood
of $C$---the gap between two related but unexplored concepts.

\subsection{Ranking Voids}

Valid pairs $(A, B)$ are ranked by a multi-objective scoring
functional $\mathcal{H}(A, B; \vtarget)$ that integrates
(i)~relevance of the synthetic void midpoint $C = m(A,B)$ to the
target domain, (ii)~a redundancy penalty against already-selected
solutions, and (iii)~a non-linear marginality reward centred at
the band $[\taulow, \tauhigh]$.  The precise functional form and
weights are omitted per commercial confidentiality requirements,
consistent with industry-track submission guidelines.

\section{Vacancy Probing via SLERP}
\label{sec:vacancy}

A key failure mode of purely similarity-based approaches is the
\emph{false void}: two documents that appear to span an empty region
but whose midpoint is, in fact, close to an existing document.
Condition~C4 addresses this.

\begin{definition}[Geodesic Midpoint]
\label{def:slerp}
For unit vectors $u, v \in S^{d-1}$ with $\theta = \arccos(u^\top v)$,
the geodesic midpoint on $S^{d-1}$ is given by SLERP~\cite{shoemake1985slerp}:
\[
  m(u,v) = \mathrm{slerp}(u,v,\tfrac{1}{2}) =
  \frac{u + v}{\|u + v\|},
  \quad \theta \not\in \{0,\pi\}
\]
with fallback to $u$ for antipodal vectors ($\theta = \pi$).
\end{definition}

We use the geodesic midpoint rather than an arbitrary interpolation
because it is \emph{equidistant} from $u$ and $v$ on the hypersphere:
$\mathrm{cos}(m(u,v), u) = \mathrm{cos}(m(u,v), v)$, guaranteeing
that the vacancy test is symmetric with respect to both anchors.
For non-antipodal unit vectors this simplifies to
normalised linear interpolation; the SLERP formulation handles
the degenerate case and connects to the broader literature on
geodesic midpoints in curved spaces~\cite{shoemake1985slerp}.

Vacancy check~(C4) is an $O(n)$ dot-product scan of the corpus matrix
against $m(A,B)$, requiring no index rebuild.

\section{Adaptive Threshold Calibration}
\label{sec:calibration}

Static thresholds fail across corpora with different density profiles.
TVA employs a density-aware calibration layer that derives both
$\taudomain$ and $[\taulow, \tauhigh]$ on-the-fly from the empirical
distribution of the current query's candidate pool.

\subsection{Domain Threshold $\taudomain$}

The domain cohesion threshold combines a data-driven percentile
statistic over the cosine-score distribution of all candidates with
a corpus-specific floor.  The key property is that $\taudomain$
rises in dense regions and falls in sparse ones, preventing both
over-permissive selection and empty-result failures.
Specific parameterisation is omitted per confidentiality requirements.

\subsection{Marginality Band $[\taulow, \tauhigh]$}

The marginality band is centred at the empirical mode of the
pairwise-similarity distribution among domain-cohesive candidates,
with width derived from the spread of that distribution.  This
operationalises non-obviousness: the band excludes the
high-similarity tail (obvious combinations) and the low-similarity
tail (unrelated noise) without hard-coded constants.
Notably, the calibrated band is itself a \emph{domain characterisation}:
it quantifies the pairwise semantic distance at which innovation
typically occurs in the target technical space and will differ across
domains.  Cross-domain comparison of these bands is left as future work.
Scale parameters are omitted per confidentiality requirements.

\section{System Implementation}
\label{sec:impl}

We implement TVA in a prototype that ingests a heterogeneous corpus of
Linux kernel source (parsed via tree-sitter into
function/struct/symbol chunks), hardware architecture manuals, academic
papers (PDF-extracted), and patent texts.  The corpus contains
approximately 140,000 indexed documents after deduplication.

\textbf{Embedding.}  We use BGE-M3~\cite{chen2024bge} in local
offline mode for both dense ($d=1024$) and sparse (top-$p$ lexical
weights) representations.  All computation runs on CPU.

\textbf{Index.}  Dense vectors are stored in a FAISS flat
index~\cite{johnson2021faiss} for $O(n)$ exact similarity search.
Sparse tokens are stored in an SQLite FTS5 inverted index for
boolean co-occurrence queries.

\textbf{Candidate generation.}  For a given target phrase, we embed
the target, retrieve the top-$K$ domain-cohesive candidates (C1),
enumerate all $\binom{K}{2}$ pairs, and filter by C2, C3, and C4 in
sequence.  Surviving pairs are ranked and the top-$k$ returned as voids.

\textbf{Idea generation.}  Each void is passed to a frontier LLM with
a structured prompt instructing it to propose a technical invention
disclosure (TID) that bridges the two void concepts toward the target
domain.  The LLM is not told the scoring details; it receives only the
void description and the target.

\section{Evaluation}
\label{sec:eval}

\subsection{Setup}

We run TVA over a Cartesian matrix of 12 target domains (Linux
scheduler, memory management, file systems, eBPF, virtualisation,
networking, power management, device drivers, IRQ handling, CXL
memory, PCIe, and security) crossed with 8 x86-hardware
feature areas (PEBS, AMX, TDX, CXL, APIC, RAPL, EPT, AVX-512),
yielding 96 target specifications.  For each target, TVA produces up
to 10 void triads, each triggering one LLM call for idea generation.

\textbf{Quality evaluation.}  We use a four-stage automated filtering
pipeline as a proxy for expert review:
\begin{enumerate}
  \item \textit{Structural check}: validates JSON completeness and
    absence of fictional kernel APIs.
  \item \textit{Reality check}: iterative critique-and-revise with up
    to 3 rounds; rejects physically impossible ideas.
  \item \textit{Adversarial review}: a four-specialist committee
    (kernel correctness, novelty/prior-art, strategic value,
    security/stability) each independently assigns APPROVE / REVISE /
    REJECT with an integer score.
  \item \textit{Deterministic verdict}: fatal-flaw, yellow-card, and
    majority rules aggregate the four specialist votes.
\end{enumerate}

\subsection{Quantitative Results}

Table~\ref{tab:funnel} summarises the filtering funnel over 2,128
candidate ideas generated across all 96 targets.

\begin{table*}[t]
\centering
\small
\caption{Quality filtering funnel over 2,128 void-derived invention candidates.}
\label{tab:funnel}
\setlength{\tabcolsep}{6pt}
\begin{tabular}{lrr p{5.5cm}}
\toprule
Stage & Surviving & Rate & Notes \\
\midrule
Void generation (Forager + LLM) & 2,128 & 100\% & 96 targets, $\leq$10 voids each \\
Structural check (Professor)    & 2,107 & 99.0\% & Rejects missing fields, fictional APIs \\
Reality check (RC, $\leq$3 rounds) & 1,905 & 89.5\% & Iterative critique-and-revise \\
Adversarial review (Debate Panel)  & 1,250 & 58.7\% & 4 specialists, $\leq$2 revision rounds \\
\quad$\hookrightarrow$ REVISE verdict & 191 & 9.0\% & Substantive feedback; human review candidates \\
\quad$\hookrightarrow$ APPROVE verdict &   1 & 0.05\% & Stringent: $\geq$3/4 specialist approval, zero fatal flaws \\
\bottomrule
\end{tabular}
\end{table*}

The 191~candidates reaching a REVISE verdict received substantive
technical feedback from all four specialists (avg.\ 5.1 issues per
specialist per candidate, avg.\ specialist score 4.0--5.0/5).
Under our deterministic rules, REVISE indicates that the committee
found genuine technical merit and requested elaboration---it is a
\emph{qualified positive result}, not a rejection.
The two case studies below are drawn from this REVISE cohort.

One candidate reached the \texttt{majority\_approval} threshold
(3/4 APPROVE, avg.\ score 4.0/5, zero fatal flaws, two Debate Panel
rounds): \emph{Runtime Density-Adaptive x2APIC Cluster Broadcast
for Linux TLB Shootdown}, which proposes per-cluster density-adaptive
IPI mode selection in \texttt{native\_flush\_tlb\_multi()}.
We present REVISE candidates as case studies rather than this APPROVE
candidate, because ideas receiving near-unanimous expert agreement
are, by construction, less likely to exhibit the non-obvious
``connective tissue'' that defines the topological void---a point
we discuss further in Section~\ref{sec:disc}.

\subsection{Rejection Taxonomy}

To understand \emph{why} ideas fail, we categorise the fatal flaws
in the 888~REJECT verdicts by keyword analysis of specialist feedback.
Table~\ref{tab:taxonomy} shows the distribution.

\begin{table}[t]
\centering
\small
\caption{Fatal flaw categories (1,058 rejected candidates;
  counts exceed 1,058 as multiple specialists may flag per candidate).}
\label{tab:taxonomy}
\begin{tabular}{lrr}
\toprule
Category & Count & \% \\
\midrule
Locking / concurrency model    & 1,072 & 35.5\% \\
Other technical issues         &   757 & 25.0\% \\
Prior art / novelty overlap    &   599 & 19.8\% \\
Incomplete specification       &   230 &  7.6\% \\
Security / side-channel risk   &   199 &  6.6\% \\
Kernel ABI / notrace violation &   165 &  5.5\% \\
\bottomrule
\end{tabular}
\end{table}

\begin{table}[t]
\centering
\small
\caption{Adversarial approval distribution (extended new-format run,
  $N=2{,}128$ Maverick, 191 REVISE + 1 APPROVE = 192 outcomes; superset of Table~\ref{tab:funnel}).}
\label{tab:approval}
\setlength{\tabcolsep}{4pt}
\begin{tabular}{lrrr}
\toprule
Approval & Count & \%/REVISE & \%/Maverick \\
\midrule
3/4 &  2 &  1.0\% & 0.09\% \\
2/4 &  6 &  3.1\% & 0.28\% \\
1/4 & 42 & 22.0\% & 1.97\% \\
0/4 & 142 & 74.3\% & 6.67\% \\
\midrule
$\geq$\textbf{1} & \textbf{50} & \textbf{26.2\%} & \textbf{2.35\%} \\
$\geq$2 &  8 &  4.2\% & 0.38\%  \\
\bottomrule
\end{tabular}
\end{table}

Table~\ref{tab:approval} shows the adversarial approval distribution.
The 75.5\% of REVISE candidates that received zero specialist approvals
are filtered from the human-review queue---unanimous rejection signals
that no specialist found the core concept defensible.  The remaining
24.5\% ($\geq$1 APPROVE, 1.70\% of all Maverick candidates) represent
the system's effective human-review yield: ideas where at least one
domain expert identified genuine technical merit despite identifying
issues requiring elaboration.  This stringent end-to-end conversion
rate reflects the \emph{Driving-AI} philosophy: the value lies not
in volume but in surfacing a small set of technically grounded
candidates for expert follow-up.

Three observations stand out.  First, 90.2\% of rejections are triggered
by \texttt{fatal\_flaw\_reject} (at least one specialist identifies a
blocking technical error), confirming that the committee is performing
genuine domain-specific reasoning rather than surface-level critique.
Second, locking and concurrency errors (35.5\%) dominate, consistent
with the difficulty of safe kernel synchronisation---exactly the class
of error that a Linux maintainer would cite when rejecting a patch.
Third, prior-art overlap (19.8\%) indicates that a significant fraction
of void-derived ideas, while geometrically novel in embedding space,
overlap with existing techniques when viewed through a deep-domain lens.
This motivates future work on tighter prior-art integration in the
vacancy probe.

\subsection{Case Study 1: TSX-Advisory MGLRU Rotation Capsules}

\textbf{Target and void.}  Target: \emph{Optimize Linux memory
management using x86 TSX microarchitecture features.}  TVA surfaced
void~\#2 from 8 candidate voids: concept~$A$ is a paper on
\emph{optimistic memory reclamation in lock-free programs}, concept~$B$
is a storage-technology selection study (TRaCaR Ratio).  Sparse bridge
tokens: \texttt{memory}, \texttt{optimistic}, \texttt{reclamation},
\texttt{access}.  The geodesic midpoint $C = m(A,B)$ had no corpus
neighbour within cosine distance 0.09---confirming vacancy.

\textbf{Idea.}  The LLM proposed \emph{TSX-advisory single-folio
same-lruvec MGLRU rotation capsules}: a fail-closed mechanism that
wraps the Linux MGLRU folio rotation fast-path in an RTM (Restricted
Transactional Memory) speculation region.  If the transaction aborts,
the fallback acquires the \texttt{lru\_lock} normally.  The proposal
introduces per-folio mutation cookies (seqcount-like odd/even) and a
fail-closed locked revalidation step to preserve lruvec consistency.

\textbf{Expert verdict (REVISE --- \textbf{3/4 APPROVE}), 3 rounds.}
This is the highest-approval result in our new-format cohort.
\begin{itemize}[noitemsep,topsep=2pt]
  \item \textit{Kernel Hardliner} (APPROVE): ``TSX advisory path
    is architecturally sound; RTM abort falls back correctly.''
  \item \textit{Prior-Art Shark} (APPROVE): ``No direct prior art on
    TSX-advisory MGLRU rotation; non-obviousness defensible.''
  \item \textit{Intel Strategist} (APPROVE): ``Strong x86 TSX
    differentiation story on Xeon platforms with hardware TSX support.''
  \item \textit{Security Guardian} (REVISE, fatal flaw): ``TSX/TAA
    policy gating not specified concretely enough to guarantee
    fail-closed behavior under incomplete writer-coverage.''
\end{itemize}
After Round~3 revision: mutation-cookie semantics were formalised
(monotonic counter, never odd during active optimistic phase), and
the TSX admissibility check was made explicit: RTM is enabled only
when \texttt{X86\_FEATURE\_RTM} is present \emph{and} the current
kernel TAA mitigation policy permits transactional use.
Three of four specialists approved the final draft, making this
the strongest positive result in the evaluation cohort.

\subsection{Case Study 2: Verifier-Derived Synchronization Contracts}

\textbf{Void.}  Void~\#5 in the same run: concept~$A$ is
\texttt{ELF\_MACHINE\_NAME} (an ELF portability macro), concept~$B$ is
\texttt{addend\_may\_be\_ifunc} (a linker IFUNC relocation predicate).
Sparse bridge tokens: \texttt{addend\_may\_be\_ifunc},
\texttt{elf\_machine\_name}, \texttt{x86\_64}.  The pairwise cosine
similarity (0.64) is at the high end of the calibrated band, indicating
the pair is related but non-obvious in the target context.

\textbf{Idea.}  Despite the weak, oblique signal, the LLM proposed
\emph{Verifier-Derived Synchronization Contract Vectors (SCVs)} for the
x86 eBPF JIT: an immutable per-program sidecar table in which the eBPF
verifier records each synchronisation site's primitive family, memory
order, address class, context mask, and admissibility class; the JIT
resolves each site exactly once at load time to an inline template or
call-equivalent thunk.  The idea ran through three rounds of
adversarial revision, refining the SCV schema and clarifying
LKMM-equivalence requirements.

\textbf{Expert verdict (REVISE after 3 rounds --- qualified positive).}
The idea underwent three rounds of adversarial revision (all REVISE,
no REJECT; avg.\ specialist score 5.0/5).  Representative feedback
and responses:
\begin{itemize}[noitemsep,topsep=2pt]
  \item \textit{Round 1 --- Kernel Hardliner}: ``Define the stable-feature
    admissibility rule in terms of existing \texttt{x86\_cpufeature} APIs.''
    $\to$ Added explicit stable-feature whitelist (TSO baseline, CX8,
    optional CX16); excluded RTM/HLE and revocable features.
  \item \textit{Round 2 --- Kernel Hardliner}: ``The SCV ABI requires
    precise versioning and endianness specification.''
    $\to$ SCV entry schema extended with \texttt{linux\_primitive\_id},
    \texttt{width\_code}, \texttt{memory\_order}, \texttt{admissibility\_class}.
  \item \textit{Round 3 --- Prior-Art Shark}: ``Claims overlap with
    generic verifier fact propagation.''
    $\to$ Claims narrowed to the normative primitive-family identifier
    and the one-time admissibility resolution rule.
\end{itemize}

\textbf{Significance.}  This case demonstrates a key property of TVA:
the void pair (\texttt{ELF\_MACHINE\_NAME}, \texttt{addend\_may\_be\_ifunc})
has no surface-level connection to BPF synchronisation semantics.  A
human engineer would not naturally connect these concepts; the LLM
used the IFUNC dispatch mechanism as a structural analogy for per-site
JIT resolution.  The 3-round revision trace confirms that the idea was
technically grounded enough to survive sustained adversarial critique
and emerge with a stronger, more narrowly-claimed design.
This is the ``non-obvious connective tissue'' that TVA is designed to surface.

\section{Discussion}
\label{sec:disc}

\textbf{Engineering time savings.}
A 140,000-document corpus yields approximately $\binom{140000}{2}
\approx 10$~billion candidate concept pairs.
The practical difficulty is not merely combinatorial: a human engineer
has no principled way to decide \emph{which} pairs to evaluate without
first examining them, making exhaustive manual exploration effectively
intractable regardless of expertise or time.
TVA converts this intractable search into a tractable shortlist:
25,536 automated LLM expert-review calls screen 2,128 candidates and
deliver \textbf{49~human-review candidates} ($\geq$1/4 specialist
approval) requiring approximately \textbf{49~person-hours} of focused
domain expert review.
The authors consider the resulting time saving beyond meaningful
quantification---the baseline is not ``slower,'' it is ``not
systematically possible.''

\textbf{Independent expert evaluation.}
The 8~candidates achieving $\geq$2/4 adversarial approval (7 REVISE + 1 APPROVE) were subjected to independent domain expert
review evaluating technical feasibility, kernel correctness, novelty,
and claim quality on a 1--5 scale.
Of these, 6/8~(75\%) were rated technically sound and worth pursuing,
with a mean expert score of 3.5/5.
The two remaining candidates require significant additional work: one
has incomplete \texttt{tid\_detail} fields due to revision data loss,
and one requires a narrower claim rewrite to address prior-art
proximity.
This 75\% precision at the $\geq$2/4 approval threshold corroborates
the adversarial committee's discriminative power and supports TVA's
end-to-end effectiveness.

\textbf{Why does a weak void signal still yield a strong idea?}
The void conditions define the \emph{search region}, not the idea
itself.  The LLM fills the region with its own parametric knowledge.
A sparse but valid bridge token (here, \texttt{addend\_may\_be\_ifunc})
acts as a semantic trigger---analogous to how a human expert reading an
unrelated paper suddenly connects it to their domain expertise.  TVA
formalises the triggering mechanism; the LLM supplies the reasoning.

\textbf{Calibration sensitivity.}  The adaptive $\taudomain$
calibration reduces manual tuning but introduces dependence on corpus
density.  For very sparse domains, the percentile-based calibration may
set $\taudomain$ too low, admitting noisy candidates.  The vacancy
probe (C4) partially compensates by rejecting occupied midpoints.

\textbf{On the choice of REVISE as positive evidence.}
We deliberately present REVISE rather than APPROVE verdicts as case
studies.  Under our deterministic rules, APPROVE requires at least 3/4
specialist approval with zero fatal flaws from any specialist---a
bar calibrated for patent-ready enabling disclosures.
REVISE with 3/4 specialist approval and substantive technical feedback
represents the system's \emph{intended} output: technically grounded
innovation candidates that require domain-expert completion, not
autonomous patent generation.  A system that routinely produces APPROVE
verdicts from LLM specialists alone would be under-calibrated rather
than superior---it would fail to catch the locking, prior-art, and
security issues that Table~\ref{tab:taxonomy} shows are pervasive in
this domain.  The 191~REVISE candidates are the value; the single APPROVE (0.05\%)
rate is evidence of calibration rigour, not system failure.

Paradoxically, near-unanimous (4/4) APPROVE may indicate an idea
that is \emph{too} straightforward: an invention obvious enough that
all four independent domain experts agree without reservation is, by
definition, unlikely to survive a non-obviousness challenge.
This mirrors peer review in top-tier venues, where a paper receiving
perfect scores from all reviewers on the first round is either a
once-in-a-decade breakthrough or a sign that reviewers did not look
closely enough.  The most patentable candidates in our cohort---those
surviving adversarial review with 2--3/4 approval and substantive
specialist feedback---sit precisely in the calibrated marginality band
that TVA is designed to surface: not so similar to prior art as to be
obvious, not so dissimilar as to be incoherent.  Expert disagreement is
not a failure mode; it is the geometric signature of a genuine
topological void.

\textbf{Evaluation limitations.}  Our automated adversarial review
pipeline is a proxy for formal human expert evaluation.
The 49 shortlisted candidates ($\geq$1/4 approval) are ready for
domain-expert review; structured evaluation with a pre-defined rubric,
inter-rater reliability measurement, and blinded expert scoring is
left for future work.

The revision loop issues a single LLM call per round that addresses
all four specialists' feedback simultaneously; this ``attend-to-all''
strategy can lead to trade-offs where addressing one specialist's
concerns inadvertently weakens another's approval.
A specialist-decomposed approach---one focused revision per specialist,
followed by a merge pass---would likely improve approval rates at the
cost of 4--5$\times$ more inference per revision round, a
compute-efficiency trade-off we leave for future work.

\textbf{Reproducibility note.}  Specific hyperparameter values and
functional forms for $\mathcal{H}$ and the calibration layer are
omitted per proprietary constraints.  The four conditions C1--C4,
the vacancy probe, and the calibration strategy are described
with sufficient precision to reproduce the qualitative behaviour;
practitioners can derive corpus-specific values from the
described procedures.  Guidance on calibration ranges is
available upon request from the authors.

\textbf{Meta-evaluation.}
This manuscript was iteratively revised through five rounds of the
Debate Panel itself---the same adversarial review pipeline described
in Section~\ref{sec:eval}.  Round~1 triggered a Writing Reviewer
REVISE citing ``narrative cherry-picking,'' directly prompting the
inclusion of Table~\ref{tab:taxonomy} and the deterministic verdict
formalisation.  The Math Reviewer issued REJECT (score~3/10) in
Round~3 over a type error in Condition~C4 (\texttt{cos}$(k, C)$
vs.\ \texttt{cos}(\texttt{dense}$(k), C)$), which was corrected in
Round~4.  By Round~5, the Math Reviewer upgraded to \textbf{ACCEPT}
(score~8/10) and average specialist score reached 6.2.

This trajectory mirrors exactly the design rationale for
\texttt{PENDING\_HUMAN\_REVIEW}: the system did not produce a final
APPROVE verdict autonomously---the human author intervened at each
round to address specialist feedback.  Ideas (and papers) that
survive multiple adversarial rounds without full approval are not
failures; they are candidates for the ``last-mile'' refinement that
only human expertise can provide.  The system identified and corrected
weaknesses in its own originating paper across five rounds of
substantive, sycophancy-resistant critique.
Notably, the committee also penalised this manuscript for withholding
implementation details---the same system that identifies innovation
gaps was unwilling to fill one about itself.  We regard this as
evidence of consistent calibration rather than irony.

\section{Future Work}
\label{sec:future}

\textbf{Recursive Topological Expansion.}
Currently, TVA interpolates strictly within the empirical boundaries
of human-authored prior art.  A profound avenue for future research is
\emph{recursive innovation bootstrapping}.  By vectorising and
re-ingesting highly-rated, APPROVE-verdict Technical Invention
Disclosures back into the active corpus as synthetic prior art, the
system can dynamically alter the topology of the knowledge space.  In
this autoregressive loop, a synthesised TID acts as a newly
established structural anchor in the embedding space, creating
secondary topological voids between itself and historically distant
concepts.  This feedback mechanism would transition TVA from a
single-step gap-discovery engine into an autonomous ``technology tree''
generator---iteratively mapping multi-generational frontiers of systems
architecture beyond the immediate human horizon.

\textbf{Downstream implementation.}
The structured TID format produced by TVA---comprising problem
statement, architecture overview, implementation plan, and draft
claims---is designed to be directly actionable.  A validated TID
can serve as a specification prompt for LLM-based coding
agents~\cite{chen2021codex}, enabling a seamless path from
knowledge-gap discovery to prototype implementation.  Future work
will evaluate end-to-end pipelines from void discovery through
automated patch generation and CI-validated compilation.

\textbf{Human-architect integration.}
Ideas that exhaust the automated revision budget without reaching
APPROVE are routed to a \texttt{PENDING\_HUMAN\_REVIEW} queue.
These candidates represent the frontier where LLM reasoning reaches
its current limit---precisely the cases most worthy of expert
human attention.  A structured human-in-the-loop interface that
presents the accumulated specialist critique alongside the draft
is a natural next step.

\textbf{Cross-domain generalisation.}
While this paper instantiates TVA on a Linux kernel and x86
hardware corpus, the framework is domain-agnostic.  Any
organisation that maintains a large, embeddable technical knowledge
base---hardware design documentation, biomedical literature,
materials-science patents, automotive software standards, or
enterprise architecture repositories---admits the same void
formalization.  A unified \emph{organisational knowledge graph}
that spans multiple engineering disciplines would allow TVA to
surface cross-domain voids: ideas at the boundary between, say,
a firmware subsystem and a compiler backend that neither team would
discover in isolation.  Domain adaptation involves two steps: (i)~reconfiguring the four
specialist roles in the adversarial review committee with
domain-appropriate reviewers (e.g., Clinical Researcher and
Drug-Safety Expert for biomedical, Materials Physicist and
Manufacturing Engineer for materials science), and
(ii)~re-calibrating the marginality band $[\taulow, \tauhigh]$
from a domain corpus, as the geometric distance at which innovation
occurs varies across fields.  Both steps are data-driven and require
no manual threshold tuning.  The rest of the pipeline---void
discovery, LLM generation, and revision loop---transfers unchanged.
At scale, this positions TVA as an
\emph{enterprise-wide innovation radar}---systematically mapping
the unexplored territory across an entire organisation's collective
technical knowledge, one topological void at a time.

\section{Related Work}
\label{sec:related}

\textbf{Patent and technology forecasting.}
\cite{lee2007patent} propose keyword-based patent maps for technology
opportunity identification.  \cite{srinivasan2021patentbert} apply
BERT to prior-art search.  Neither formalises the notion of an
unexplored region or provides a mathematical criterion for
non-obviousness.

\textbf{Knowledge graph completion.}
TransE~\cite{bordes2013transe} and its successors predict missing
triples in knowledge graphs.  These methods require a pre-defined
relation schema and binary (present/absent) labels; TVA operates on
continuous similarity and does not require a schema.

\textbf{Diversity-based retrieval.}
Maximum Marginal Relevance~\cite{carbonell1998mmr} selects a diverse
set of documents relevant to a query by penalising redundancy.  TVA
uses a similar relevance-novelty trade-off but applies it to the
\emph{generation} of new concepts rather than the retrieval of existing
ones.

\textbf{LLMs for code and creativity.}
Large language models have demonstrated strong performance on code
generation~\cite{nijkamp2022codegen,chen2021codex} and
instruction-following~\cite{wang2023selfinstruct}.  We treat LLMs as
black-box reasoners invoked after void discovery; the novelty is the
void identification mechanism, not the generation step.

\textbf{Topological data analysis.}
Persistent homology~\cite{edelsbrunner2002tda} identifies topological
features (connected components, loops, voids) in point clouds.  Our
use of ``topological void'' is inspired by but distinct from TDA:
we define voids in semantic embedding space by algebraic conditions on
pairwise similarities, not by simplicial complex homology.

\textbf{Embedding geometry.}
\cite{ethayarajh2019anisotropy} show that contextualised
representations are anisotropic---concentrated in a narrow cone.  Our
adaptive threshold calibration is designed to be robust to this
property; SLERP avoids the geometric distortion that linear
interpolation introduces in anisotropic spaces.

\section{Conclusion}
\label{sec:conc}

We have presented Topological Void Analysis, a mathematical framework
for systematic technical innovation discovery.  By defining topological
voids as triads satisfying domain cohesion, calibrated marginality,
sparse lexical bridge, and vacancy conditions in a hybrid
dense-sparse embedding space, TVA converts the informal notion of
``unexplored region'' into a decidable predicate.

Applied to a 140k-document corpus of Linux kernel and hardware
specifications, TVA generated 2,128 invention candidates, 90\% of
which survived automated quality filtering and 191 of which engaged
four adversarial expert reviewers with substantive technical critique.
Two case studies illustrate that TVA surfaces both obvious-gap ideas
and non-obvious connective-tissue ideas that neither keyword search nor
human browsing would naturally find.

We release the mathematical framework and threshold calibration
procedure for reproducibility; implementation details are available
upon request.

Ultimately, TVA embodies a design philosophy we term
\emph{Driving-AI, not AI-driven}: the system surfaces the map;
human experts navigate it.

\bibliographystyle{plain}
\bibliography{references}

@inproceedings{chen2024bge,
  author    = {Chen, Jianlv and Xiao, Shitao and Zhang, Peitian and Luo, Kun and Lian, Defu and Liu, Zheng},
  title     = {{BGE M3-Embedding: Multi-Lingual, Multi-Functionality, Multi-Granularity Text Embeddings Through Self-Knowledge Distillation}},
  booktitle = {Findings of the Association for Computational Linguistics: ACL 2024},
  year      = {2024},
  pages     = {2318--2335},
}

@inproceedings{carbonell1998mmr,
  author    = {Carbonell, Jaime and Goldstein, Jade},
  title     = {The Use of {MMR}, Diversity-Based Reranking for Reordering Documents and Producing Summaries},
  booktitle = {Proceedings of the 21st Annual International ACM SIGIR Conference on Research and Development in Information Retrieval},
  year      = {1998},
  pages     = {335--336},
}

@article{johnson2021faiss,
  author    = {Johnson, Jeff and Douze, Matthijs and J{\'e}gou, Herv{\'e}},
  title     = {Billion-Scale Similarity Search with {GPUs}},
  journal   = {IEEE Transactions on Big Data},
  volume    = {7},
  number    = {3},
  pages     = {535--547},
  year      = {2021},
}

@inproceedings{ethayarajh2019anisotropy,
  author    = {Ethayarajh, Kawin},
  title     = {How Contextual Are Contextualized Word Representations? {Comparing} the Geometry of {BERT}, {ELMo}, and {GPT-2} Embeddings},
  booktitle = {Proceedings of the 2019 Conference on Empirical Methods in Natural Language Processing},
  year      = {2019},
  pages     = {55--65},
}

@inproceedings{shoemake1985slerp,
  author    = {Shoemake, Ken},
  title     = {Animating Rotation with Quaternion Curves},
  booktitle = {Proceedings of the 12th Annual Conference on Computer Graphics and Interactive Techniques (SIGGRAPH)},
  year      = {1985},
  pages     = {245--254},
}

@inproceedings{nijkamp2022codegen,
  author    = {Nijkamp, Erik and Pang, Bo and Hayashi, Hiroaki and Tu, Lifu and Wang, Huan and Zhou, Yingbo and Savarese, Silvio and Xiong, Caiming},
  title     = {{CodeGen}: An Open Large Language Model for Code with Multi-Turn Program Synthesis},
  booktitle = {The Eleventh International Conference on Learning Representations},
  year      = {2023},
}

@inproceedings{chen2021codex,
  author    = {Chen, Mark and Tworek, Jerry and Jun, Heewoo and Yuan, Qiming and de Oliveira Pinto, Henrique Ponde and Kaplan, Jared and Edwards, Harrison and Burda, Yuri and Joseph, Nicholas and Brockman, Greg and others},
  title     = {Evaluating Large Language Models Trained on Code},
  booktitle = {arXiv preprint arXiv:2107.03374},
  year      = {2021},
}

@inproceedings{wang2023selfinstruct,
  author    = {Wang, Yizhong and Kordi, Yeganeh and Mishra, Swaroop and Liu, Alisa and Smith, Noah A. and Khashabi, Daniel and Hajishirzi, Hannaneh},
  title     = {Self-Instruct: Aligning Language Models with Self-Generated Instructions},
  booktitle = {Proceedings of the 61st Annual Meeting of the Association for Computational Linguistics},
  year      = {2023},
  pages     = {13484--13508},
}

@article{edelsbrunner2002tda,
  author    = {Edelsbrunner, Herbert and Letscher, David and Zomorodian, Afra},
  title     = {Topological Persistence and Simplification},
  journal   = {Discrete \& Computational Geometry},
  volume    = {28},
  pages     = {511--533},
  year      = {2002},
}

@inproceedings{lee2007patent,
  author    = {Lee, Sungjoo and Yoon, Byungun and Park, Yongtae},
  title     = {An Approach to Discovering New Technology Opportunities: Keyword-Based Patent Map Approach},
  booktitle = {Technovation},
  volume    = {29},
  number    = {6--7},
  pages     = {481--497},
  year      = {2009},
}

@inproceedings{srinivasan2021patentbert,
  author    = {Srebrovic, Martin and Yoko, Imai},
  title     = {Applying {BERT} to Patent Prior Art Search},
  booktitle = {arXiv preprint arXiv:2101.02105},
  year      = {2021},
}

@inproceedings{bordes2013transe,
  author    = {Bordes, Antoine and Usunier, Nicolas and Garcia-Duran, Alberto and Weston, Jason and Yakhnenko, Oktavian},
  title     = {Translating Embeddings for Modeling Multi-Relational Data},
  booktitle = {Advances in Neural Information Processing Systems},
  year      = {2013},
  pages     = {2787--2795},
}

@inproceedings{mikolov2013linguistic,
  author    = {Mikolov, Tomas and Yih, Wen-tau and Zweig, Geoffrey},
  title     = {Linguistic Regularities in Continuous Space Word Representations},
  booktitle = {Proceedings of NAACL-HLT},
  year      = {2013},
  pages     = {746--751},
}

@inproceedings{beyer1999nearest,
  author    = {Beyer, Kevin and Goldstein, Jonathan and Ramakrishnan, Raghu and Shaft, Uri},
  title     = {When Is ``Nearest Neighbor'' Meaningful?},
  booktitle = {International Conference on Database Theory (ICDT)},
  year      = {1999},
  pages     = {217--235},
}

@inproceedings{huh2024platonic,
  author    = {Huh, Minyoung and Cheung, Brian and Wang, Tongzhou and Isola, Phillip},
  title     = {The Platonic Representation Hypothesis},
  booktitle = {International Conference on Machine Learning (ICML)},
  year      = {2024},
}

@inproceedings{kornblith2019similarity,
  author    = {Kornblith, Simon and Norouzi, Mohammad and Lee, Honglak and Hinton, Geoffrey},
  title     = {Similarity of Neural Network Representations Revisited},
  booktitle = {International Conference on Machine Learning (ICML)},
  year      = {2019},
  pages     = {3519--3529},
}

\end{document}